\documentclass[sort&compress,final]{aipproc}
\layoutstyle{8x11single}
\usepackage{amsmath}
\usepackage{amssymb}
\usepackage{bm}
\usepackage{booktabs}
\begin{document}
\title{Advances in QCD sum-rule calculations} 
\author{Dmitri Melikhov}{address={Institute for High
Energy Physics, Austrian Academy of Sciences, Nikolsdorfergasse
18, A-1050 Vienna, Austria},
altaddress={D.~V.~Skobeltsyn Institute
of Nuclear Physics, M.~V.~Lomonosov Moscow State University, Moscow, Russia}}
\begin{abstract}
We review the recent progress in the applications of QCD sum rules to hadron properties  
with the emphasis on the following selected problems: 
(i) development of new algorithms for the extraction of ground-state parameters from two-point correlators; 
(ii) form factors at large momentum transfers from three-point vacuum correlation functions;  
(iii) properties of exotic tetraquark hadrons from correlation functions of four-quark currents. 

\end{abstract}
\maketitle
 
\section{Introduction}
The method of sum rules is 35 years old. In spite of this respectable age, 
the method is being permanently enriched by new ideas and new calculations  
and remains one of the widely used and competitive tools both for the determinations 
of the fundamental QCD parameters (e.g., quark masses and $\alpha_s$) and for the 
calculation of hadron properties. In this talk we review the recent progress in the 
applications of QCD sum rules to hadron properties with the emphasis on the selected topics: 
(i) sum rules for two-point vacuum correlation functions and leptonic decay constants of heavy mesons;
(ii) sum rules for three-point vacuum correlation functions, form factors and three-meson couplings; 
(iii) sum rules for exotic tetraquark states. 

QCD sum rules \cite{svz} (see also \cite{colangelo,ioffe} for further references) 
is one of the main {\it analytic} methods for 
the study of hadron properties from the field-theoretic Green functions (correlators) in full QCD.  
The correlators are calculated by means of the Wilsonian operator product expansion (OPE) which provides the rigorous framework for the separation of long and short distances, in QCD being dominated by 
nonperturbative and perturbative physics, respectively \cite{nsvz1984}. The OPE clearly identifies, e.g., the origin 
of chiral symmetry breaking and 
the emergence of hadron masses, leads to factorization of complicated amplitudes of hadron interactions 
at large momentum transfers. 

\noindent$\bullet$ 
QCD sum rules provide hadron amplitudes which satisfy all rigorous properties imposed by perturbative QCD and, 
at the same time, contain nonperturbative contributions determined in a unique way. 
As an OPE-based method, QCD sum rules are formulated in the Euclidean region. 
However, by combining OPE with the knowledge of the analytic structure of the Green functions and resummation schemes, 
the analytic continuation to the Minkowski space may be performed. In this respect QCD sum rules may have 
a broader range of applicability than lattice QCD. Last but not least, as an analytic method, QCD sum rules provide physics 
insights in the hadron structure, which are not easy to get from the numerical results of lattice QCD. 

\noindent$\bullet$ 
The method of QCD sum rules favourably compares with other analytic methods, such as effective theories or functional 
methods: the method of sum rules is based on the Wilsonian OPE in full QCD and therefore involves no 
other implicit assumptions often present in other analytic method. 

\vspace{.1cm}
\noindent {\bf a. OPE and the sum rule for the correlator} 

\noindent The basic object in the method of QCD sum rules -- as well as in lattice QCD -- is the vacuum-to-vacuum correlator, 
i.e., the vacuum average of the $T$-product of quark and gluon currents. In lattice QCD, one finds this correlator 
numerically at large values of the Euclidean time $\tau$. In the method of QCD sum rules, one calculates the correlator 
analytically as the Taylor expansion in $\tau$. Technically, one considers a so-called Borelized correlator, 
i.e. applies the Borel transform to the Feynman diagrams, written as spectral representations in the energy variables. 
The inverse Borel mass parameter is related to $\tau$. The OPE provides the analytic double expansion of this correlator 
in form of a perturbatively calculable power series 
in the strong coupling constant $\alpha_s$ and in powers of $\tau$; the ``power corrections'' --- terms 
involving powers of~$\tau$ --- 
are given via {\it condensates}, 
expectation values of gauge-invariant operators over the physical vacuum in QCD; these condensates describe in an unambiguous 
way nonperturbative QCD contributions. 

Alternatively, one may derive a representation for the Borelized correlator in terms of the
intermediate hadron states. The two representations for the Borelized correlator 
--- by OPE and by sum over hadron states --- constitute the two sides of the QCD sum~rule.

\vspace{.1cm}

\noindent {\bf b. Isolating the ground-state contribution from the Borelized correlator}

\noindent At large $\tau$, the ground-state dominates the correlator which thus fully determines the ground-state parameters. 
In the region of small and intermediate $\tau$, where the truncated OPE gives a good description of the correlator, 
excited hadronic states give sizeable contributions. 
In order to get rid of the excited states and to isolate the ground-state contribution from the correlator, one invokes 
the idea~of quark--hadron duality \cite{shifman1,shifman2,lm}: the excited states are dual to high-energy parts of Feynman 
diagrams of perturbative QCD. The ground-state contribution is then equal to the ``{\it dual correlator}'' -- 
the correlator in which the spectral integrals for perturbation theory diagrams 
are cut at a certain {\it effective continuum threshold} $s_{\rm eff}$, or simply ``effective threshold''.  
The effective continuum threshold differs from the physical continuum threshold determined by masses of low-lying hadrons. 
Obviously, apart from a truncated OPE for a correlator, the effective continuum threshold 
is a crucial ingredient of every sum-rule extraction of ground-state parameters;
this quantity governs the accuracy of the quark--hadron duality and determines to large extent the numerical 
value of the extracted parameters of the bound state. The truncated OPE itself cannot provide precise values  
of the ground-state parameters. Therefore, the method of QCD sum rules provides hadron parameters with some uncertainty 
which is referred to as systematic uncertainty \cite{lms_2ptsr}. 
 
{\it Understanding the properties of the effective continuum threshold  
and finding a criterion for fixing this quantity is the key to obtaining reliable hadron parameters from sum rules}.

\section{1. Two-point correlation function and the OPE}
Let us start with the simplest object -- the two-point correlation function; the perturbative expansion for this object 
is known to a higher accuracy compared to more complicated correlators. Because of that, the formulation and application 
of the appropriate and reliable algorithms for the extraction of the hadron parameters from this correlator is 
becoming increasingly important. 

The two-point function, i.e. the vacuum average of the $T$-product of two interpolating quark currents 
is the basic object for the sum-rule calculation of the decay constants of the heavy-light mesons 
such as $B$, $B_s$, $D$, $D_s$ or their vector analogues. For instance, for heavy-light pseudoscalar currents 
$j_5=\overline{m}_b\bar qi\gamma_5b$ (here $\overline{m}_b$ is the scale-dependent $\overline{\rm MS}$ mass of the heavy quark 
and $M_b$ will denote its pole mass; the light-quark mass is neglected) one obtains 
\begin{eqnarray}
\Pi(p^2)=i\int
d^4x\,e^{ipx}\left\langle\Omega \left|T\left(j_5(x)j^\dag_5(0)\right)\right|\Omega\right\rangle
\end{eqnarray}
The Wilson OPE for the $T$-product and for the correlation function has the following form: 
\begin{eqnarray}
T\left(j_5(x)j_5^\dagger(0)\right)=C_0(x^2,\mu)\hat 1 + \sum\limits_n C_n(x^2,\mu) :\hat O(x=0,\mu):
\end{eqnarray}
and 
\begin{eqnarray}
\Pi(p^2)=\Pi_{\rm pert}(p^2,\mu)+\sum_n \frac{C_n}{(p^2-M_b^2)^n}\langle \Omega| :\hat O(x=0,\mu): |\Omega \rangle
\end{eqnarray}
Here the physical QCD vacuum $|\Omega\rangle$ is a complicated object which differs from perturbative 
QCD vacuum $|0\rangle$. The properties of the physical vacuum are characterized by the 
{condensates} -- the nonzero expectation values of gauge-invariant operators over this physical vacuum: 
\begin{eqnarray}
\langle \Omega|:\hat O(0,\mu):| \Omega\rangle\ne 0.
\end{eqnarray}
The numerical estimates for the condensates may be found in \cite{colangelo,ioffe}. 
Here we only list the recent determinations of the lowest-dimension condensates which claim an extremely high accuracy: 
\begin{eqnarray}
\langle \Omega | \bar q q(2\;{\rm  GeV}| \Omega\rangle^{\overline{\rm MS}}=(282\pm 2\;{\rm MeV})^3 \;
[9], \qquad 
\langle \Omega |\frac{\alpha_s}{\pi} G^{a}_{\mu\nu}G^{a, \mu\nu}| \Omega\rangle=0.013\pm 0.0016\;{\rm GeV^4} \; 
[10]. 
\end{eqnarray}
The two-point function satisfies the dispersion representation (which requires subtractions not shown here) 
\begin{eqnarray}
\Pi(p^2)=\int \frac{ds}{s-p^2}\rho(s), 
\end{eqnarray}
and may be calculated both using OPE (which gives it in the form $\Pi_{\rm OPE}(p^2)$) 
and using the sum over the hadron intermediate states 
(which gives it in the form $\Pi_{\rm hadr}(p^2)$). The sum rule is the statement that both forms 
represent the same quantity and thus should be equal to each other 
\begin{eqnarray}
\label{srp2}
\Pi_{\rm OPE}(p^2)=\Pi_{\rm hadr}(p^2). 
\end{eqnarray}
The spectral densities for the two representations read 
\begin{eqnarray}
\rho_{\rm OPE}(s)=\left[
\rho_{\rm pert}(s,\mu)+\sum_n C_n \delta^{(n)}(s-M_b^2)\langle\Omega|O_n(\mu)|\Omega\rangle
\right], \qquad 
\rho_{\rm hadr}(s)=f^2_BM_B^4\delta(s-M_B^2)+\rho_{\rm cont}(s)
\end{eqnarray}
Here $M_B$ denotes the heavy-meson mass, $f_B$ is its decay constant defined as 
\begin{eqnarray}
\langle0|j_5|B\rangle=f_BM_B^2. 
\end{eqnarray}
The truncated OPE series has quark and gluon singularities and does not have the hadron ones; 
therefore, comparison of the truncated OPE and the hadron representation in (\ref{srp2}) may be done in the region of $p^2$ far from 
hadron thresholds and resonances. 

Performing the Borel transform which serves several purposes (suppressing the contribution of 
the excited states, killing the subtraction terms in the dispersion representation for $\Pi(p^2)$, 
improving the convergence of the perturbative expansion \cite{svz}) one arrives at the Borel image 
of the two-point function 
\begin{eqnarray}
\Pi_{\rm hadr}(\tau)=\int ds \exp(-s \tau)\rho_{\rm hadr}(s)=f_B^2 M_B^4 e^{-M_B^2\tau}+
\int\limits_{s_{\rm phys}}^\infty ds\,e^{-s\tau}\rho_{\rm hadr}(s), 
\end{eqnarray}
where $s_{\rm phys}=(M_{B^*}+M_P)^2$ is the physical continuum threshold, determined by the masses of hadrons 
which may appear as the intermediate states, and 
\begin{eqnarray}
\Pi_{\rm OPE}(\tau)=\int ds \exp(-s \tau)\rho_{\rm OPE}(s)=
\int\limits_{m_b^2}^\infty ds\,e^{-s\tau}\rho_{\rm pert}(s,\mu)+\Pi_{\rm power}(\tau,\mu),  
\end{eqnarray}
where power corrections $\Pi_{\rm power}(\tau,\mu)$ are given via the condensates and radiative corrections to them. 

The sum rule now takes the form 
\begin{eqnarray}
\Pi_{\rm OPE}(\tau)=\Pi_{\rm hadr}(\tau). 
\end{eqnarray}
Recall that the hadron (i.e. full-QCD) representation $\Pi_{\rm hadr}(\tau)$ is an infinite sum of the exponential terms, whereas 
power corrections in $\Pi_{\rm OPE}(\tau)$ contain polynomials in $\tau$ multiplied by $\exp(-M_b^2\tau)$. 
Therefore the truncated OPE provides a good 
description of $\Pi_{\rm hadr}(\tau)$ at ``not too large'' values of $\tau$. This determines the choice of the {\it Borel window} -- 
the working $\tau$-range where the OPE gives an accurate description~of the
exact correlator (i.e., all higher-order radiative and power
corrections are under control) and at the same time the ground state gives
a ``sizable'' contribution to the correlator. 

The best-known 3-loop calculations of the perturbative 
spectral density \cite{chetyrkin} have been performed in form of
an expansion in terms of the $\overline{\rm MS}$ strong coupling
$\alpha_{\rm s}(\mu)$ and the pole mass $M_b$:
\begin{eqnarray}
\rho_{\rm pert}(s,\mu)
=\rho^{(0)}(s,M_b^2)+\frac{\alpha_{\rm s}(\mu)}{\pi}\rho^{(1)}(s,M_b^2)+
\left(\frac{\alpha_{\rm s}(\mu)}{\pi}\right)^2\rho^{(2)}(s,M_b^2,\mu)+\cdots.
\end{eqnarray}
An alternative option \cite{jamin} is to reorganize the perturbative expansion in terms of the running $\overline{\rm MS}$
mass, $\overline{m}_b(\nu)$,~by substituting $M_b$ in the spectral
densities $\rho^{(i)}(s,M_b^2)$ via its perturbative expansion in
terms of the running mass $\overline{m}_b(\nu)$ 
\begin{eqnarray}
M_b=\overline{m}_b(\nu)\left(1+\frac{\alpha_s(\nu)}{\pi}\,r_1
+\left(\frac{\alpha_s(\nu)}{\pi}\right)^2r_2+\ldots \right).
\end{eqnarray}
As noticed in \cite{jamin,hoang}, two different scales, $\mu$ and $\nu$, naturally emerge when
reorganizing the perturbative expansion from the pole $b$-quark
mass to the running $b$-quark mass. In our discussion we do not distinguish between these scales, but in practical 
calculations the scales have been treated independently.

\subsection{Advanced algorithms for an isolation of the ground-state contribution}

The hadron representation contains the sum over all hadron intermediate states, whereas we are primarily interested in the 
ground state contribution. To exclude the excited-state contributions, one adopts the {\it duality Ansatz}: all contributions of
excited states are counterbalanced by the perturbative contribution above an {\em effective continuum threshold},~
$s_{\rm eff}(\tau,\mu)$ which differs from the physical continuum threshold. 
Applying the duality assumption yields:
\begin{eqnarray}
\label{duality}
f_B^2 M_B^4 e^{-M_B^2\tau}=\int\limits_{m_b^2}^{s_{\rm eff}(\tau,\mu)}ds\,e^{-s\tau}\rho_{\rm pert}(s,\mu)+
\Pi_{\rm power}(\tau,\mu)\equiv\Pi_{\rm dual}(\tau,s_{\rm eff}(\tau,\mu)).
\end{eqnarray}
The rhs is the {\em dual correlator} $\Pi_{\rm dual}(\tau,s_{\rm eff}(\tau))$ (we shall not explicitly write $\mu$ as an argument 
of $s_{\rm eff}$ but this dependence should be kept in mind). 
{\it Obviously, even if the QCD inputs $\rho_{\rm pert}(s,\mu)$ and $\Pi_{\rm power}(\tau,\mu)$ are known,  
the extraction of the decay constant requires $s_{\rm eff}(\tau,\mu)$.} 
Let us emphasize, that the effective threshold should be the function of $\tau$ and $\mu$: 
(i) one can easily check that $s_{\rm eff}$ should depend on $\tau$ in order the $\tau$-dependences of the r.h.s. and the 
l.h.s. of (\ref{duality}) match each other; (ii) since the truncated OPE is used in the r.h.s. of (\ref{duality}), 
the effective threshold also depends on the choice of the scale $\mu$. 

In early applications of the method of sum rules, it was common to use the approximation $s_{\rm eff}(\tau)=const$; the 
value of this constant has been fixed by requiring the maximal stability 
(i.e. the least unphysical dependence of the hadron observable on the Borel parameter $\tau$). This procedure 
proved to work reasonably well, although it did not allow one to probe the uncertainty of the extracted hadron parameter 
induced by using the approximation of a constant effective continuum threshold. 

It should be emphasized that even if the OPE for the correlation function is known with very high accuracy in the Borel window, 
the hadron parameters can still be determined with some uncertainty which reflects the limited intrinsic accuracy of the method 
of sum rules. We refer to the 
corresponding uncertainty as to the {\em systematic uncertainty}. The latter is related to the adopted prescription 
for fixing~the effective continuum threshold $s_{\rm eff}(\tau)$. 

As the accuracy of the OPE for the correlation functions has increased, one faced the acute necessity 
to provide more accurate and reliable procedures for the extraction of hadron parameters: gaining control over the systematic 
uncertainties has become mandatory \cite{lms_2ptsr}.  

The results of \cite{lms_new} demonstrated that in those cases where the bound-state mass $M_B$ is known, one can use it and 
improve the accuracy of the decay constant. 
We introduce the {\em dual invariant mass\/} $M_{\rm dual}$ and the {\em dual decay constant\/} $f_{\rm dual}$
\begin{eqnarray}
M_{\rm dual}^2(\tau)\equiv-\frac{d}{d\tau}\log\Pi_{\rm dual}(\tau,s_{\rm eff}(\tau)),
\qquad f_{\rm dual}^2(\tau)\equiv M_B^{-4}\,e^{M_B^2\tau}\,\Pi_{\rm dual}(\tau,s_{\rm eff}(\tau)).
\end{eqnarray}
The deviation of $M_{\rm dual}(\tau)$ from~$M_B$ measures the contamination of the dual correlator by excited states. 

Starting with any trial function for $s_{\rm eff}(\tau)$ and minimizing the deviation of $M_{\rm dual}$ from $M_B$ in
the $\tau$-window yields a variational solution for $s_{\rm eff}(\tau)$. As soon as the latter is found, one readily 
obtains the corresponding decay constant $f_B$ from (\ref{duality}). 

We consider polynomials in $\tau$ and obtain   
their parameters by minimizing the squared difference between $M^2_{\rm dual}$ and $M^2_B$ in the $\tau$-window:
\begin{eqnarray}
\chi^2\equiv\frac{1}{N}\sum_{i=1}^N\left[M^2_{\rm dual}(\tau_i)-M_B^2\right]^2.
\end{eqnarray}
As shown in several exactly solvable models, the band of the estiamtes for $f_B$ corresponding to the variational solutions for 
linear, quadratic,~and cubic trial $s_{\rm eff}(\tau)$, provides a realistic estimate~for the systematic uncertainty 
of the decay constant \cite{qcdvsqm,lms2011jpg}. 

The resulting $f_{B}$ obtained according to the procedure described above is sensitive to the input values of all the OPE parameters 
(quark masses, $\alpha_s$, the condensates) which are known with some uncertainties thus yielding 
the {\em OPE-related uncertainty\/} of $f_B$. 
To obtain the latter, one assumes the Gaussian distributions~for the OPE parameters mentioned above. 
Moreover, because of the truncation of the OPE series, the decay constants exhibit an unphysical dependence on the precise value 
of the renormalization scales $\mu$. A priori, any choice of the scale is equivalently good; therefore, 
we average over the scale in some intervals assuming the {\it uniform} distribution of $\mu$.

\noindent $\bullet$ 
Another simple algorithm for fixing the $\tau$-dependent effective threshold in the Borel sum rule 
has been recently adopted in \cite{khodj}: for each value of $\tau$ the authors calculated $M_{\rm dual}(\tau)$ {\it neglecting} 
the $\tau$-dependence of $s_{\rm eff}(\tau)$ and then easily obtain $s_{\rm eff}$ by solving the equation $M_{\rm dual}(\tau)=M_B$. 
Obviously, the resulting effective thresholds do depend on $\tau$; neglecting their $\tau$-dependence while calculating 
the dual mass leads to some intrinsic inconsistencies. Following our old idea, we  
tested the algorithm of \cite{khodj} in a quantum-mechanical potential model for the case of a potential which contains the 
confining and the Coulomb parts \cite{qcdvsqm}. This analysis shows that in quantum mechanics the algorithm with the variational 
solutions described above provides more 
reliable and accurate estimates for the decay constants of the heavy-light mesons compared with the algorithm of \cite{khodj}.

\noindent $\bullet$
An interesting approach to the extraction of the ground-state parameters within the finite-energy sum rule has been formulated and applied to 
the decay constants of heavy-light and heavy-heavy mesons in \cite{dominguez}. We have also tested this algorithm in the potential 
model \cite{qcdvsqm}. 
For the potential-model parameters appropriate for for heavy-light mesons the algorithm of \cite{dominguez} was shown to provide 
rather accurate estimates for the decay constants such that the ``invisible'' systematic error remains at a few percent level only.  


\subsection{Charm sector}
\begin{figure}[b]
\centering
\begin{tabular}{cc}
\includegraphics[width=6.3cm]{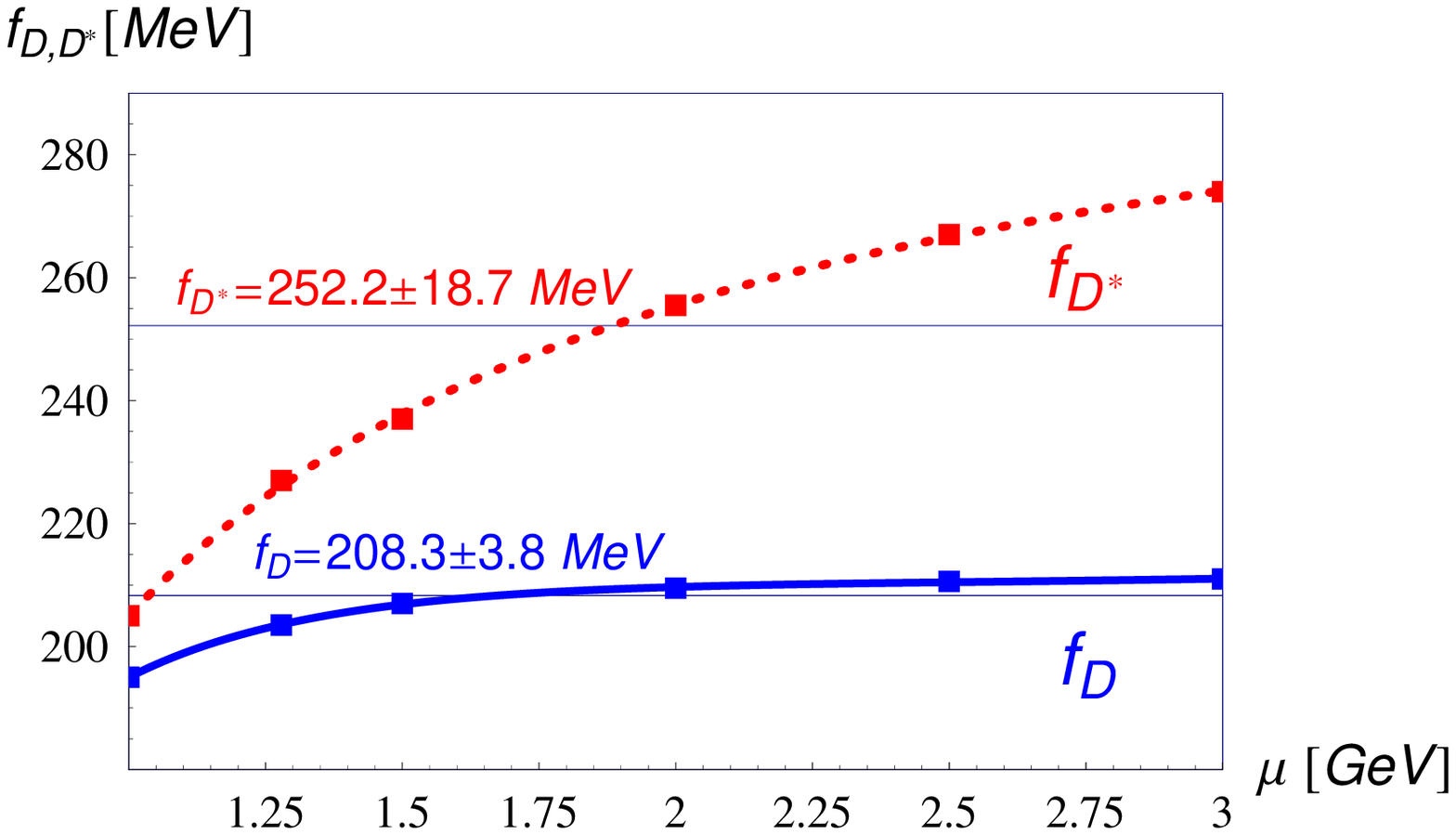} &
\includegraphics[width=6.3cm]{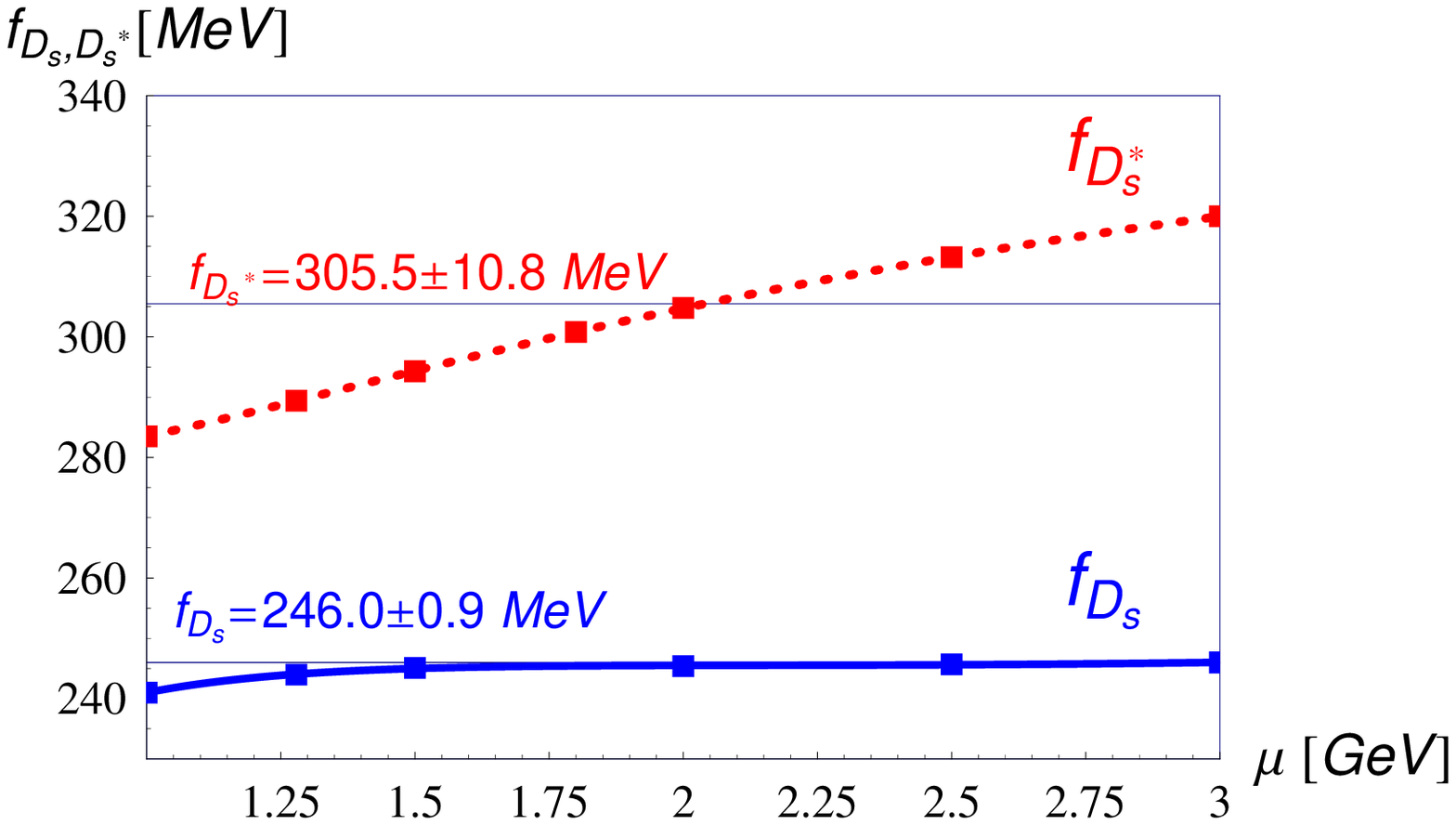} 
\end{tabular}
\caption{Decay constants of $D$, $D_s$ $D^*$ and $D_s^*$ mesons depending on the scale $\mu$. 
\label{Plot:1a}}
\end{figure}

For the extraction of the decay constants of the charmed pseudoscalar and vector mesons, one makes use of the best-known three-loop expression 
for the spectral densities of the two-point functions for pseudoscalar and vector currents. The OPE in terms of the 
pole mass $M_b$ calculated in \cite{chetyrkin} does not exhibit a perturbative hierarchy, therefore one rearrange the OPE in terms of the 
running $\overline{\rm MS}$-mass \cite{jamin}. Then, the perturbative hierarchy of the correlation function starts to 
depend on $\mu$; this feature allows one to choose the range of $\mu$ where the perturbative hierarchy is visible. 
The negative effect of this rearrangement of the perturbative expansion is that,  
because of the truncation of the OPE series, the extracted decay constants acquire an unphysical dependence on the scale $\mu$. 
In the charm sector this however does not lead to any serious problems. Figure \ref{Plot:1a} shows the dependence of the decay constants of 
the charmed pseudoscalar and vector mesons for the central values of all other OPE parameters after applying the algorithm for fixing the 
effective thresholds described above. One can see a weak $\mu$-dependence of the decay constants of the pseudoscalar mesons mesons, whereas for 
vector mesons this $\mu$-dependence is more pronounced. Averaging over the OPE parameters in their respective intervals and over the scale in the 
range $1\le \mu[GeV]\le 3$ one arrives at the following results \cite{lms_charm} 
\begin{eqnarray}
&&f_{D}=(208.3 \pm 7.3_{\rm OPE}\pm 5_{\rm syst})\; {\rm MeV}, \qquad 
f_{D_s}=(246.0 \pm 15.7_{\rm OPE}\pm 5_{\rm syst})\; {\rm MeV}\nonumber\\
&&f_{D^*}=(252.2 \pm 22.3_{\rm OPE}\pm 4_{\rm syst})\; {\rm MeV}, \qquad 
f_{D_s^*}=(305.5 \pm 26.8_{\rm OPE}\pm 5_{\rm syst})\; {\rm MeV}.
\end{eqnarray}
For the ratio we reported $f_{D^*}/f_D= 1.221\pm 0.080_{\rm OPE}\pm 0.008_{\rm syst}$, which compares nicely with the lattice QCD result 
$f_{D^*}/f_D= 1.20\pm 0.02$. The results for the charmed mesons from other sum-rule analyses \cite{khodj,narisonfB} 
agree well with each other and with the results from lattice QCD \cite{lms_b2014}. 

\subsection{Beauty sector}
Similar to the charm sector, the OPE for pseudoscalar and vector currents containing the $b$-quark, does not show any perturbative hierarchy; 
there is no reason to assume that the unknown higher-order perturbative contributions are small. Rearranging the perturbative expansion in 
terms of the running mass introduces the dependence of the scale $\mu$ and opens the possibility to choose the working range of $\mu$ 
in which the perturbative hierarchy is explicit thus allowing to hope the unknown higher orders do not contribute 
substantially to the correlation function. 

In the $b$-sector one encounters two interesting features of the sum-rule analysis:

\noindent $\bullet$ 
The sum-rule results for the beauty-meson decay constants correlate very strongly with the $b$-quark mass \cite{lms_bmass2013}
\begin{eqnarray}
{\delta f_B}/{f_B}\approx-8\,{\delta m_b}/{m_b}, 
\end{eqnarray}
$m_b\equiv \overline{m}_b(\overline{m}_b)$. Making use of $m_b=4.18\pm 0.03$ GeV \cite{pdg} leads to $f_B>210$ MeV, in clear tention with the recent lattice QCD results 
for $f_B\sim 190$ MeV. Combining our sum-rule analysis with $f_B$ and $f_{B_s}$ from lattice QCD~yields \cite{lms_bmass2013}
\begin{eqnarray}
m_b=(4.247\pm0.027\pm 0.011_{\rm(syst)})\;{\rm GeV}. 
\end{eqnarray}
The sum-rule results for the decay constants corresponding to this value of the $b$-quark mass read   
\begin{eqnarray}
f_{B}=(192.0 \pm 14.3_{\rm OPE}\pm 3.0_{\rm syst})\; {\rm MeV},\qquad 
f_{B_s}=(228.0 \pm 19.4_{\rm OPE}\pm 4.0_{\rm syst})\; {\rm MeV}
\end{eqnarray}
$\bullet$ 
For the decay constant of $B^*$, one observes an unexpectedly strong $\mu$-dependence \cite{lms_bstar2014}:
\begin{figure}[t]
\centering
\begin{tabular}{cc}
\includegraphics[width=6.3cm]{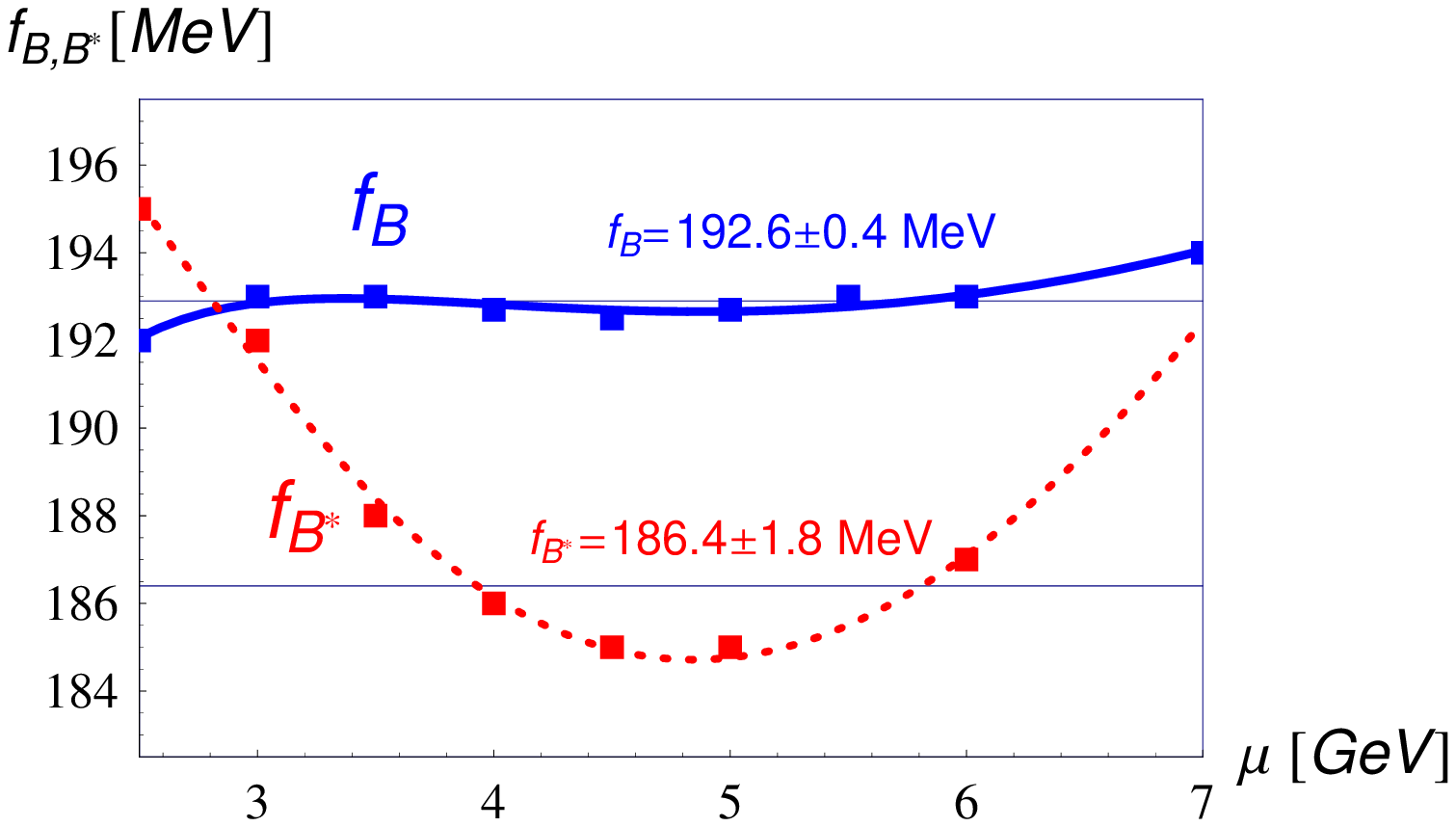} & \includegraphics[width=6.3cm]{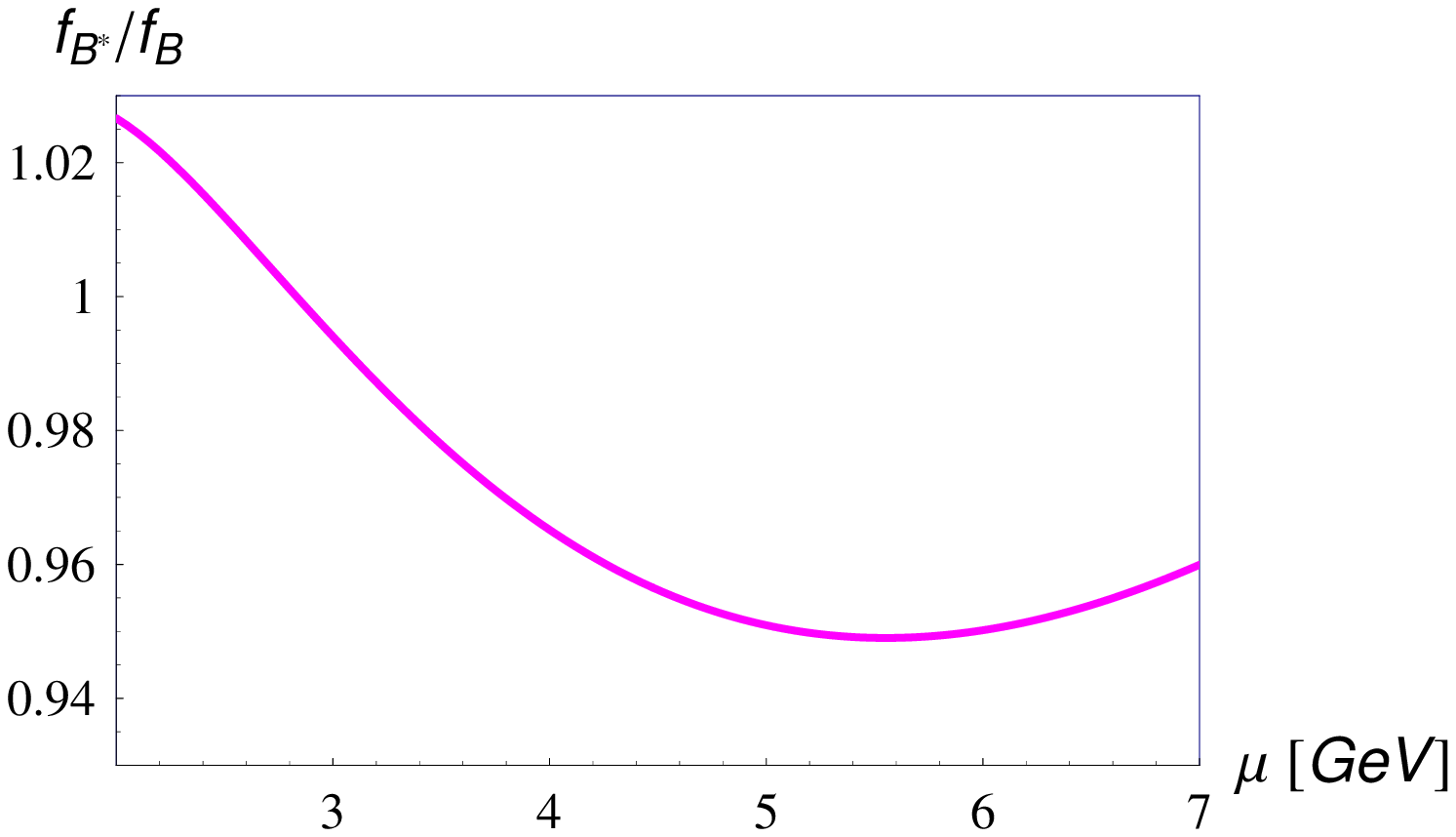} 
\end{tabular}
\caption{Decay constants of $B$ and $B^*$ mesons and the ratio $f_{B^*}/f_B$ depending on the scale $\mu$. 
\label{Plot:2}}
\end{figure}
Averaging over the scale range $3 < \mu [{\rm GeV}]< 6$ leads to 
$$
f_{B^*}/f_B=0.923\pm 0.059, \qquad f_{B_s^*}/f_{B_s}=0.932\pm 0.047. 
$$
Taking into account only low-scale results for $2.5 < \mu [{\rm GeV}]< 3.5$, 
yields $f_{B^*}/f_B=0.994\pm 0.01$. 
The sum-rule analysis \cite{khodj} also gives indications that $f_{B^*}/f_B \le 1$ (ses Table II of \cite{khodj}). 
Surprisingly, the QCD sum-rule prediction for $f_{B^*}/f_B$ is below the
corresponding results from lattice QCD, which~seem to favour a value slightly above unity \cite{lms_b2014,becirevic}. 
Clearly, such tension calls for further detailed investigations.


\subsection{$\mu$-dependence of the physical quantities}

The heavy-light correltors are known with an impressive three-loop accuracy and are therefore rather weakly sensitive to 
the variations of the scale. Nevertheless, the {\it dual correlator} of the vector currents which includes the low-energy region of the 
Feynman diagrams only and, respectively, the vector-meson decay constants are rather sensitive to the choice of the scale. 
In many cases this scale-dependence is the main sourse of the OPE uncertainty in the decay constants. 
We should mention that in some publications the $\mu$-dependence is treated in a specific way \cite{narisonfB}: one just chooses one 
scale at which the decay constant has, e.g., an extremum in $\mu$, and provides the results for this very scale assigning no theoretical 
uncertainty to the scale fixing. This of course reduces strongly the total uncertainty of the decay constant obtained with the 
sum-rule technique but 
from our point of view such a treatment is not justified: the (unphysical) $\mu$-dependence is an effect of the truncation of 
the OPE series and thus reflects an essential feature of QCD. Any of the scale for which a reasonable perturbative 
hierarchy is seen, may be used for the determination of the hadron parameter; the unpleasant $\mu$-dependence of the sum-rule 
results should 
be thus properly reflected in the theoretical uncertainty of the hadron parameter obtained using a QCD sum rule.


\section{Sum rules for three-point vacuum correlation functions}
Let us now discuss the calculation of the meson elastic and transition form factors from the three-point vacuum 
correlation functions \cite{ioffe3pt,radyushkin}. The basic object in this case has the form 
\begin{eqnarray}
\label{3pt}
\Gamma(p^2,p'^2,q^2)=\int \langle \Omega|T (j(x) j(0) j(y))|\Omega \rangle \exp(-i px)\exp(-i p'y)dx dy.
\end{eqnarray}
The three-point Green function in full QCD contains the double pole related to the mesons in the $p^2$ and $p'^2$-channels in the timelike region. 
The residue in this double pole is the form factor of interest. 
The Green function in the spacelike region may be calculated using the same method as the two-point function, i.e. by performing the OPE. 
One represents the Green function $\Gamma(p^2,p'^2,q^2)$ as a double spectral integral in $p^2$ and $p'^2$, 
performs the double Borel transform $p^2\to \tau$ and $p'^2\to \tau'$ (which, similar to the two-point function, kills the subtraction 
terms and suppresses the contributions of the excited states), equate to each other the OPE and the hadron representations for 
$\Gamma(p^2,p'^2,q^2)$, and use duality property to isolate the ground-state contribution, thus 
relating the meson form factor to the low-energy region of the triangle diagrams of perturbative QCD and power corrections given through 
the condensates. For instance, the pion elastic form factor, in which case one sets $\tau=\tau'$, has the form \cite{radyushkin} 
\begin{eqnarray}
\label{duality3pt}
F_\pi(Q^2)\,f_\pi^2=\int\limits_0^{s_{\rm eff}(Q^2,\tau)}\,\int\limits_0^{s_{\rm eff}(Q^2,\tau)}
{\rm d}s_1\,{\rm d}s_2\, \Delta_{\rm pert}(s_1,s_2,Q^2)
e^{-\frac{s_1+s_2}{2}\tau}
+\frac{\left<\alpha_s\,G^2\right>}{24\pi}\,\tau
+\frac{4\pi\,\alpha_s\left<\bar qq\right>^2}{81}\,\tau^2
\left(13+Q^2\,\tau\right)+\cdots,
\nonumber \\
\Delta_{\rm pert}(s_1,s_2,Q^2)=\Delta^{(0)}(s_1,s_2,Q^2)+\alpha_s \Delta^{(1)}(s_1,s_2,Q^2)+\cdots.\hspace{4cm}
\end{eqnarray}
An essential feature of the three-point sum rule is that the effective threshold now depends on the Borel parameter $\tau$ and 
the momentum transfer $Q$ \cite{lm2012prd,m_lcsr,hagop}; obviously, one faces a serious problem of finding appropriate algorithms 
for fixing $s_{\rm eff}(Q^2,\tau)$. It should be understood that the effective 
continuum threshold for the form factor differs from the effective threshold for the decay constant.\footnote{The effective thresholds for the 
baryon form factors are strongly sensitive to the choice of the interpolating current for a specific baryon.} 

For large $Q^2$, power corrections calculated in terms of the local condensates rise as polynomials with $Q^2$, thus 
preventing a direct use of the sum rule (\ref{duality3pt}) at large $Q^2$. 
There are essentially only two possibilities to study the region of large $Q^2$ starting with the vacuum correlators: 

$\bullet$ use nonlocal condensates which are aimed at the resummation of the local condensate effects \cite{bakulev,bakulev2}.  

$\bullet$ work in the so-called local-duality (LD) limit $\tau=0$ \cite{bakulev}. A specific feature of this limit is that 
all power corrections vanish in this limit and details of non-perturbative dynamics are hidden 
in one complicated object -- the $Q^2$-dependent effective threshold $s_{\rm eff}(Q^2)$. 

A similar treatment may be performed for, e.g., the $\pi^0\to \gamma\gamma^*$ transition form factor \cite{lm2012jpg,ms2012prd,teryaev}
for which one obtains the single spectral representation in the LD limit: 
\begin{eqnarray}
F_{\pi\gamma}(Q^2)\,f_\pi=\int\limits_0^{\bar s_{\rm eff}(Q^2)}\,{\rm d}s\ \sigma_{\rm pert}(s,Q^2)
\end{eqnarray}
Due to properties of the spectral functions $\Delta_{\rm pert}(s_1,s_2,Q^2)$ and $\sigma_{\rm pert}(s,Q^2)$, the form factors obey the factorization theorems
\begin{eqnarray}
F_{\pi}(Q^2) \to  8\pi\alpha_s(Q^2)f_\pi^2/Q^2,\qquad F_{\pi\gamma}(Q^2)\to  \sqrt{2} f_\pi/Q^2, \qquad  f_\pi=130 \mbox{ MeV} 
\end{eqnarray}
as soon as the effective thresholds satisfy  
\begin{eqnarray}
s_{\rm eff}(Q^2\to\infty)=\bar s_{\rm eff}(Q^2\to\infty)={4\pi^2f_\pi^2}. 
\end{eqnarray}
Remarkably, due to the QCD factorization theorems for the hard form factors, the effective thresholds at $Q^2\to\infty$ are given through the 
decay constants of the participating mesons. It should be emphasized that the only feature of theory relevant for this property 
of $s_{\rm eff}(Q^2)$ is {\it factorization} of hard form factors.  

For finite $Q^2$, the effective thresholds $s_{\rm eff}(Q^2)$ and $\bar s_{\rm eff}(Q^2)$ depend on $Q^2$ 
and differ from each other \cite{blm2008plb,blm2012prd}. 
Nevertheless, setting $s_{rm eff}(Q^2)=s_{rm eff}(Q^2\to\infty)$ for all not too small $Q^2$ \cite{radyushkin}  
provides an approximate {\it parameter-free} prediction for the form factors which is becoming increasingly accurate as soon as $Q^2$ increases. 
The results of \cite{blm2012prd} give convincing evidences that $s_{\rm eff}(Q^2)$ and $\bar s_{\rm eff}(Q^2)$ are close to their asymptotic values 
already at relatively low values $Q^2\approx 4-8$ GeV$^2$. 

Thus, the LD approximation for the form factors---which requires as its crucial ingredient the knowledge of $O(1)$ and $O(\alpha_s)$ double 
spectral densities---is increasingly accurate in the region not too close to zero recoil. 
The LD approximation is very promising for the application to, e.g., heavy-to-light weak form factors. 
A still missing ingredient here is the two-loop $O(\alpha_s)$ double spectral density of the triangle diagram for different 
currents and arbitrary quark masses in the loop. This is a really challenging calculation which however opens the possibilities of very 
interesting applications. So far the only known results correspond to all massless quarks in the loop \cite{braguta} and to HQET \cite{bagan,fulvia}.

\section{Sum rules for the exotic polyquark currents}
The OPE for the correlation functions of the exotic polyquark currents involving 4 (or 5) quark fields 
of the type 
\begin{eqnarray}
D(x)=\bar q_1(x)\hat O q_2(x) \bar q_3(x)\hat O q_4(x)
\end{eqnarray}
where $\hat O$ is an appropriate combination of the Dirac matrices and possibly also of the (covariant) derivatives, 
have specific features compared to the OPE for the bilinear currents of the form $j(x)=\bar q_1(x)\hat O q_2(x)$ used 
for usual ``nonexotic'' mesons. Namely, the lowest-order $O(1)$ contribution to the OPE for any correlator involving the exotic 
current, e.g. $\Pi_{DD}=\langle 0|T(D(x)D(0)|0\rangle$, is given by the disconnected diagrams. As known from the general features 
of the Bethe-Salpeter equation and also emphasized 
recently by Weinberg \cite{Weinberg}, these disconnected diagrams are not related to the exotic bound states. 
The {\it connected} diagrams relevant for the exotic states emerge in the OPE for any correlator at the order $O(\alpha_s)$ 
and higher; therefore for the analysis of the exotic states the knowledge of the radiative corrections is mandatory. 
This makes the analysis of the exotic states a more technically involved problem than the analysis of the normal hadrons. 

Nevertheless, due to the fact that the observed exotic states are narrow, the procedure of extracting their parameters 
from the OPE has the same features and the same challenges as for the normal hadrons. Our experience in the analysis of the 
usual hadrons proves that a truncated OPE for the correlation function does not allow one to study at the same time both 
the {\it existence} of the isolated ground state and of its {\it properties}. However, if the mass of the narrow bound state is known, 
the method of sum rules allows one to obtain reliable predictions for its decay constants and the form factors. 

\subsection{Structure of the exotic tetraquark states}
Obviously, the exotic tetraquark states may have a rather complicated ``internal'' structure; 
two most popular scenarios of this structure are  
a confined tetraquark state (i.e. a bound state in a confining potential between the two color-triplet diquarks) and a molecular 
``nuclear-physics like'' bound state in the system of two colorless mesons. 

However, an important question about the structure of the exotic state---which to large extent determines also its production 
mechanism---is not easy to answer \cite{jaffe}: 
(i) by a combined color-spinor Fierz rearrangement of the tetraquark interpolating current $D(x)$  
one can write it either in diquark-antidiquark or meson-meson form; 
(ii) the same quantum numbers of the exotic interpolating current may be obtained by different combinations of its diquark-antidiquark 
or meson-meson bilinear parts. 

The simplest characteristic of a usual meson is its decay constant, i.e. the transition amplitude between the vacuum and the meson 
induced by its interpolating current; for a heavy quarkonium state the decay constant is analogous to its wave function 
at the origin $\psi(r=0)$. 

For an exotic tetraquark state one should considers the {\it connected} self-energy functions 
\begin{eqnarray}
\Pi_{DD}=\langle 0|T(D(x)D(0)|0\rangle \equiv \langle DD\rangle
\end{eqnarray}
and study the corresponding sum rules. However, for an exotic state one may obtain a set of the decay constants, related to 
different structure of the interpolating current with the quantum numbers of the exotic tetraquark of interest. 
{\it The answer to the question of the dominant structure of the tetraquark may be given only by the analysis of a large 
set of the decay constants.}

\noindent $\bullet$ As the first step, one needs to study systematically the interpolating currents for tetraquark currents with 
different quantum nembers. As the next step one can calculate the set of $\Pi_{DD}$. 
Because of the factorization property of the two-point function of the local tetraquark currents \cite{narison}, 
the radiative corrections to $\Pi_{DD}$ are given via radiative corrections to the various two-point functions of 
the bilinear quark currents. For some of these two-point functions (namely, $\langle VV\rangle$ and 
$\langle AA\rangle$) the radiative corrections are well-known, for some of them (such as $\langle TT\rangle$, $T$ is the tensor 
bilinear current) these corrections should be calculated. 

\noindent $\bullet$ Then, the set of the sum rules for different two-point functions $\Pi_{DD}$ should be studied and only then 
the answer about the structure of the observed narrow exotic candidates may be obtained. 
Especially interesting cases here are the narrow charged tetraquark $Z^-(4430)$ ($J^P=1^+$ 
and the width $\simeq 45$ MeV, valence-quark content $\bar  cc \bar u d$) and 
X(3872) ($J^{PC}=1^{++}$ X(3872), the width $<24$ MeV). 

Another interesting possibility---so far not discussed in the literature---is considering {\it nonlocal} interpolating currents for 
the exotic mesons. The nonlocality of the interpolating currents should allow one to access in a better way 
subtle details of the tetraquark structure.    

\subsection{Strong fall-apart decays of the exotic tetraquark states}
\begin{figure}[b]
\label{Fig:exotic}
\includegraphics[width=12cm]{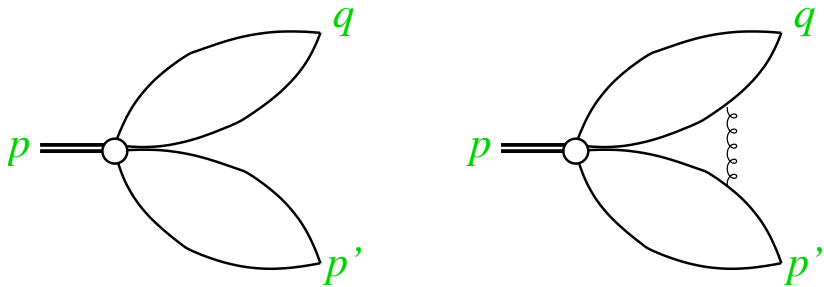}
\caption{
(a) The disconnected $O(1)$ diagram which does not depend on the variable $p^2$ relevant for the tetraquark properties;
(b) One of the lowest-order connected $O(\alpha_s)$ diagram which contributes to the tetraquark decay amplitude. }
\end{figure}

In the last decade, QCD sum rules have been extensively applied to the analysis of strong decays of exotic multiquark states 
(see e.g. \cite{nielsen2010,nielsen2014} and references therein). 
The basic object for the analysis of these decays in QCD is the three-point functions of the type 
\begin{eqnarray}
\Gamma(p,p',q)=\int\langle 0|T(D(0)j(x_1)j(x_2)|0\rangle\exp(-i p'x_1 -i q\,x_2)dx_1 dx_2.
\end{eqnarray}
This correlator contains the triple-pole in the Minkowski region
\begin{eqnarray}
\Gamma_{\rm hadr}(p,p',q)=\frac{f_X f_{M_1} f_{M_2}g_{X M_1 M_2}}{(p^2-M_X^2)({p'}^2-M_1^2)(q^2-M_2^2)}+\cdots 
\end{eqnarray}
where dots stay for less singular terms. Here $g_{X M_1 M_2}$ is the three-hadron coupling which describes the $X\to M_1M_2$ 
transition; $f_X$, $f_{M_1}$, and $f_{M_2}$ are the decay constants of the mesons describing the strength of their 
their interaction with the interpolating current 
$\langle X|D(0)|0\rangle=f_{X}$ and $\langle M_{1,2}|j_{1,2}(0)|0\rangle=f_{1,2}$ 
(we omit here all Lorentz indices and for simplicity neglect the spins of the hadrons and of the interpolating currents). 
The OPE allows one to calculate the expansion of this correlator at 
the spacelike momenta far from the hadron thresholds. Again, the leading contribution in $\alpha_s$ 
is given by a disconnected diagram 
(see Fig.\ref{Fig:exotic}a) which factorizes and does not depend on the momentum of the exotic current $p^2$ at all: 
\begin{eqnarray}
\Gamma_{\rm OPE}(p^2,p'^2,q^2)=\Pi(p'^2)\Pi(q^2)+\alpha_s\Gamma_{\rm connected}(p^2,p'^2,q^2)
\end{eqnarray} 
 
Performing the Borel transform $p^2\to \tau$, which comprises one of the steps of the sum-rule analysis, 
we see that the Borel image of the disconnected leading-order contribution vanishes.\footnote{We emphasize 
that for the decay of a usual hadron, the $O(1)$ contribution is given by a triangle diagram which depends on all three variables 
$p^2, p'^2,q^2$ and therefore of course does not vanish under the Borel transform $p^2\to \tau$; for the decays of the usual hadrons 
the $O(1)$ contribution of the perturbative QCD indeed provides the dominant contribution to the decay of interest.} 
Therefore any attempt to extract the tetraquark decay amplitude from the leading-order contribution is inconsistent. 
Relevant for the exotic-state properties are the $O(\alpha_s)$ corrections which are technically very difficult. 
This is a difficult calculation but it should be done before one may hope to get 
reliable predictions for the tetraquark properties. So far these corrections have been calculated only for the three-point 
function of the bilinear currents in two cases 
(i) for massless quarks and (ii) for infinitely heavy active quark and a massless spectator. 
For the $O(\alpha_s)$ corrections to the three-point functions $\Gamma$, involving one tetraquark and two bilinear currents,  
no results exist in the literature. 

Nevertheless, the common feature of {\it all} previous calculations of these decays within QCD sum rules 
(e.g. \cite{nielsen2014,wang}) was the attempt to study the tetraquark (and pentaquark) 
decays basing on the factorizable leading-order contribution which intrinsically has no relationship with the tetraquark 
properties (which is clear both from the factorization property $\Gamma(p,p',q)=\Pi(p'^2)\Pi(q^2)$ 
and from the large-$N_c$ behaviour of the QCD diagrams emphasized by Weinberg \cite{Weinberg}. 
Therefore the existing analyses should be strongly revised by calculating and taking into account the nonfactorizable 
two-loop $O(\alpha_s)$ corrections.

Nonzero results based on the leading-order correlation function may be obtained only by a trick. Let us consider e.g. the decay 
$Z\to \psi' + \pi^-$. One makes use of the tetraquark current
$j(x)=\bar c(x) c(x)\bar u(x)d(x)$ (we again omit the Dirac matrices for simplicity). 
The corresponding three-point correlation function of interest is 
\begin{eqnarray}
\Gamma(p^2,p'^2,q^2)&=&\int d^4x d^4y 
\exp(-i p' x)\exp(-i qy)
\langle 0| T(\bar c(0) c(0)\bar u(0) d(0), \bar c(x) c(x), \bar d(y) u(y))|0\rangle. 
\end{eqnarray}
A nonzero result for the Borel transform of the disconnected zero-order contribution may be 
obtained by first considering the soft-pion limit $q\to 0$, i.e. $p'=p$, which gives for the disconnected contribution 
$\Pi(p^2)\Pi(0)$ and then performing the Borel transform $p^2\to\tau$. However, the decay rate obtained in this way is not 
really trustworthy. 

We therefore conclude that {\it the ``fall-apart'' decay mechanism of exotic hadrons differs from the decay mechanism 
of the ordinary hadrons and requires the appropriate treatment within QCD sum rules. The calculation of the 
radiative corrections is mandatory for a reliable analysis of the properties of the exotic states.}

\section{Summary and Outlook}
In the recent years, great progress has been seen both in the calculations of the OPE 
series for various correlation functions and in the direction of formulating advanced algorithms for the extraction of 
the individual hadron parameters from these correlators. We could not discuss all the developments in this talk but let us try 
to mention in this summary the interesting open issues to be addressed in the future analyses:   

\vspace{.2cm}
$\bullet$
Let us recall that combining moment QCD sum rules with experimental/lattice data gives the most accurate estimates of 
the heavy-quark masses \cite{erler}. 

\vspace{.2cm}
$\bullet$
{\it Hadron properties from 2-point functions:}
 
\begin{itemize}
\item[a.]
We have seen a visible progress in developing the new algorithms for extracting ground state parameters from the OPE of the correlators
and gaining control over the systematic errors of the decay constants (finite-energy sum rules, Borel sum rules). 
Although it seems impossible to predict both masses and decay constants with a controlled accuracy, using the mass 
of the ground state as input, systematics can be controlled). 

\item[b.]
We have encountered interesting puzzles in the $b$-sector: 

(i) The $b$-quark mass $4.18$ GeV \cite{pdg} when used in the Borel sum rules for $f_B$ leads to tension with lattice results for $f_B$. 

(ii) Unexpectedly strong scale-dependence of decay constants of vector mesons and of $f_{B^*}/f_B$ even using the $O(\alpha_s^2)$ 
correlation function. 

\item[c.]
Calculation of the decay constants of heavy-quarkonium states within the method of QCD sum rules is still not fully settled: 
The problem here is that the OPE for the doubly-heavy correlation functions contain {\it relatively small} nonperturbative 
power corrections. Therefore in QCD, the structure of OPE for the heavy-quarkonium system is somewhat similar to the structure of OPE for a 
purely coulomb system. Obviously, the algorithms adopted and tested for light or heavy-light hadrons in which cases the nonperturbative 
corrections are large, may work differently for heavy quarkonium states. This feature may be the origin of the tensions between the sum-rule 
predictions and the results from lattice QCD and other nonperturbative approaches for e.g. the decay constants of $B_c$ mesons and some 
charmonium states \cite{narisonfB,becirevic2}. A more critical analysis of the procedures of an isolation of the ground-state 
contribution from the correlation function and in particular of the way of obtaining the systematic uncertainties is necessary. 

\item[d.]
Since the accuracy of the isolation of the ground-state contribution from the correlation function can be controlled, 
one may try to apply the method for the analysis of the excited states. Very little efforts in this direction have been 
done so far. 
\end{itemize}

\vspace{.1cm}
$\bullet$
{\it Meson elastic and transition form factors from three-point functions}

The Borel sum rules at $\tau=0$ (the so-called local duality limit) open an interesting possibility of obtaining parameter-free 
predictions for the elastic and the transition form factors of light mesons in a broad range of the momentum transfers. 
The crucial ingredients necessary for these calculations are the $O(1)$ and $O(\alpha_s)$ spectral densities of the 
triangle diagrams. As soon as these are known, the effective thresholds are determined in a unique way by the QCD 
factorization theorems for hard form factors. Assuming the effective thresholds to weakly depend on the momentum transfers---a 
hypothesis which finds support in the data for the pion form factors---one obtains the parameter-free predictions 
for the form factors in a broad range of the momentum transfers. Our analysis suggests that these representations for the 
form factors work with a few percent accuracy for $Q^2\ge$ a few GeV$^2$. 
It seems very promising to apply the same ideas to heavy-to-light transition form factors. The main problem here is 
the necessity to calculate the radiative corrections to the triangle diagrams which is a very difficult task which needs 
serious efforts. As soon as this ambitious task is fulfilled, QCD sum rules could provide parameter-free predictions for the form 
factors, increasingly accurate with increase of $Q^2$. 

\vspace{.2cm}
$\bullet$ 
{\it Baryon elastic and transition form factors} 

The calculations for baryons are obviously technically extremely involved. Many sum-rule analyses of the baryon elastic 
and transition form factors have been presented in the recent years (see e.g. \cite{anikin,aliev1,aliev2,altug} and references therein).  
With a few exceptions (e.g. \cite{anikin}), these calculations are based on the leading-order $O(1)$ correlation 
functions and use the traditional approaches to fix the effective thresholds, usually neglecting the $\tau$- and $Q^2$-dependence 
of the latter. These analyses are expected to provide reasonable ball-park estimates for the form factors; however, in most of 
the cases, the estimates of the OPE-errors (related to the uncertainties of the QCD parameters, to the missing radiative corrections, 
and in particular to a strong dependence on the scale $\mu$) and the systematic errors, related to the adopted procedures 
of fixing the effective continuum thresholds) are not done properly. Many efforts are still to be done in the domain of 
the baryon form factors.

\vspace{.2cm}
$\bullet$
{\it Three-meson strong couplings of the type $g_{D^*D\pi}$}

These quantities have been extensively addressed using three-point vacuum correlators and the corresponding sum rules. 
Again, the radiative corrections to the correlation functions have not been taken into account. Moreover, the results for 
the decay constants require extrapolations over large ranges of the momentum transfers. Therefore, one cannot expect 
good accuracy of these estimates. In many cases, the results from sum rules lead to an unrealistic picture of the $SU(3)$-breaking effects (see \cite{bracco} and refs therein). For a real progress, one needs the calculation and the inclusion of 
the radiative corrections to the appropriate three-point functions. 

\vspace{.2cm}
$\bullet$
{\it Properties of the exotic tetraquark states}

The ``fall-apart'' decay mechanism of exotic hadrons differs from the decay mechanism 
of the ordinary hadrons and requires the appropriate treatment within QCD sum rules. In distinction to the 
decays of the usual hadrons, where the knowledge of the radiative corrections is necessary for improving the accuracy 
of the sum-rule form factor calculations, the calculation of the 
radiative corrections is {\it mandatory} for a reliable analysis of the properties of the exotic states. 
The decays of the exotic states are intrinsically unrelated to the $O(1)$ disconnected correlation functions;  
the results obtained from these O(1) correlators cannot be treated as fundamental and reliable.   

\vspace{.2cm}
From this summary of the recent advances and still open issues it seems obvious that the future progress in the sum-rule 
calculations of the properties of the usual and the exotic hadrons will be related (i) to the calculations of the radiative 
corrections to the correlation functions and (ii) to further development of the appropriate algorithms for the extraction 
of the properties of the individual hadrons from these correlators.

\begin{theacknowledgments}
I have pleasure to thank Wolfgang Lucha and Silvano Simula for a pleasant and fruitful  
collaboration on the topics discussed in this talk. Many thanks are due to the Organizers for maintaining a  
stimulating and friendly atmosphere of this Conference for already more than a decade.  
\end{theacknowledgments}

\bibliographystyle{aipproc}

\begin{thebibliography}{99}
\bibitem{svz} 
M.~A.~Shifman, A.~I.~Vainshtein, and V.~I.~Zakharov, 
\emph{Nucl.~Phys.~B} {\bf 147}, 385 (1979).

\bibitem{colangelo}
P.~Colangelo and A.~Khodjamirian, 
hep-ph/0010175. 

\bibitem{ioffe}
B.~L.~Ioffe, 
\emph{Prog.~Part.~Nucl.~Phys.~} {\bf 56}, 232 (2006). 

\bibitem{nsvz1984}
V.~A.~Novikov, M.~A.~Shifman, A.~I.~Vainshtein, and V.~I.~Zakharov,
\emph{Nucl.\ Phys.\ B} {\bf 249}, 445 (1985). 

\bibitem{shifman1} 
B. Blok, M. Shifman, and D.-X. Zhang,
\emph{Phys. Rev. D} {\bf 57}, 2691 (1998). 
\bibitem{shifman2}
M. Shifman, 
hep-ph/0009131. 

\bibitem{lm}
W. Lucha and D. Melikhov, 
\emph{Phys. Rev. D} {\bf 73}, 054009 (2006).

\bibitem{lms_2ptsr}
W.~Lucha, D.~Melikhov, and S.~Simula, 
\emph{Phys.~Rev.~D} {\bf 76}, 036002 (2007); 
\emph{Phys.~Lett.~B}{\bf 657}, 148 (2007). 

\bibitem{condensate_qq}
C. T. H. Davies, {\it et al.}, 
arXiv:1301.7204. 

\bibitem{condensate_GG}
C.~Dominguez, L.~Hernandez, and K.~Schilcher, 
arXiv:1411.4500. 

\bibitem{chetyrkin}
K.~G.~Chetyrkin and M.~Steinhauser, 
\emph{Phys.~Lett.~B} \textbf{502}, 104 (2001);  
\emph{Eur.~Phys.~J.~C} \textbf{21}, 319 (2001). 

\bibitem{jamin}
M.~Jamin and B.~Lange, 
\emph{Phys.~Rev.~D} {\bf 65}, 056005 (2002). 

\bibitem{hoang}
B.~Dehnadi, A.~Hoang, V.~Mateu, and S.M.~Zebarjad, 
\emph{JHEP} {\bf 1309}, 103 (2013).

\bibitem{lms_new}
W.~Lucha, D.~Melikhov, and S.~Simula, 
\emph{Phys.~Lett.~B} {\bf 671}, 445 (2009); 
\emph{Phys.~Rev.~D} {\bf 79}, 096011 (2009);
\emph{J.~Phys.~G} {\bf 37} (2010). 

\bibitem{qcdvsqm}
W.~Lucha, D.~Melikhov, and S.~Simula, 
\emph{Phys.~Lett.~B} {\bf 687}, 48 (2010). 

\bibitem{lms2011jpg}
W.~Lucha, D.~Melikhov, and S.~Simula,
\emph{J.~Phys.~G} {\bf 38}, 105002 (2011). 

\bibitem{khodj}
P.~Gelhausen, A.~Khodjamirian, A.~A.~Pivovarov, and D.~Rosenthal, 
\emph{Phys.~Rev.~D\/} \textbf{88} (2013) 014015; \textbf{89} (2014) 099901(E). 

\bibitem{dominguez}
M.~J.~Baker, J.~Bordes, C.~A.~Dominguez, J.~Penarrocha, and K.~Schilcher, 
\emph{JHEP} {\bf 1407}, 032 (2014).




\bibitem{lms_charm}
W.~Lucha, D.~Melikhov, and S.~Simula,
\emph{Phys.~Lett.~B} {\bf 701}, 82 (2011); 
\emph{Phys.~Lett.~B} {\bf 735}, 12 (2014). 


\bibitem{narisonfB}
S.~Narison, 
arXiv:1404:6642[hep-ph].

\bibitem{lms_b2014}
W.~Lucha, D.~Melikhov, and S.~Simula,
arXiv:1411.3890. 

\bibitem{lms_bmass2013}
W.~Lucha, D.~Melikhov, and S.~Simula,
\emph{Phys.~Rev.~D} {\bf 88}, 056011 (2013). 

\bibitem{pdg} 
K.~A.~Olive {\it  et al.} (Particle Data Group), \emph{Chin.~Phys.~C} {\bf 38}, 090001 (2014). 

\bibitem{lms_bstar2014}
W.~Lucha, D.~Melikhov, and S.~Simula,
arXiv:1410.6684

\bibitem{becirevic}
D.~Becirevic, A.~Le Yaouanc, A.~Oyanguren, P.~Roudeau, and F.~Sanfilippo, 
arXiv:1407.1019 [hep-ph].


\bibitem{ioffe3pt} 
B.~L.~Ioffe and A.~V.~Smilga, 
\emph{Phys.~Lett.~B} {\bf 114}, 353 (1982).

\bibitem{radyushkin} 
V.~A.~Nesterenko and A.~V.~Radyushkin, 
\emph{Phys.~Lett.~B} {\bf 115}, 410 (1982).

\bibitem{lm2012prd}
W.~Lucha and D.~Melikhov, 
\emph{Phys.~Rev.~D} {\bf 86}, 016001 (2012). 
\bibitem{m_lcsr}
D.~Melikhov, 
\emph{Phys.\ Lett.~B} {\bf 671}, 450 (2009).
\bibitem{hagop}
W.~Lucha, D.~Melikhov, H.~Sazdjian, and S.~Simula, 
\emph{Phys.~Rev.~D} {\bf 80}, 114028 (2009). 

\bibitem{bakulev}
A.~P.~Bakulev and A.~V.~Radyushkin, 
\emph{Phys.~Lett.~B} {\bf 271}, 223 (1991).
\bibitem{bakulev2}
A.~P.~Bakulev, A.~V.~Pimikov, N.~G.~Stefanis, 
\emph{Phys.~Rev.~D} {\bf 79}, 093010 (2009). 

\bibitem{lm2012jpg}
W.~Lucha and D.~Melikhov,
\emph{J.~Phys.~G} {\bf 39}, 045003 (2012).
\bibitem{ms2012prd}
D.~Melikhov and B.~Stech, 
\emph{Phys.~Rev.~D} {\bf 85}, 051901 (2012); 
\emph{Phys.~Lett.~B} {\bf 718}, 488 (2012). 
\bibitem{teryaev}
Y.~N.~Klopot, A.~G.~Oganesian, and O.~V.~Teryaev, 
\emph{Phys.~Rev.~D} {\bf 84}, 051901 (2011); 
\emph{Phys.~Lett.~B} {\bf 695}, 130 (2011). 

\bibitem{blm2008plb}
V. Braguta, W. Lucha, and D. Melikhov, 
\emph{Phys.~Lett.~B} {\bf 661}, 354 (2008).
\bibitem{blm2012prd}
I.~Balakireva, W.~Lucha and D.~Melikhov,
\emph{J.~Phys.~G} {\bf 39}, 055007 (2012);  
\emph{Phys.~Rev.~D} {\bf 85}, 036006 (2012).



\bibitem{braguta}
V.~Braguta and A.~Onishchenko,  	
\emph{Phys.~Lett.~B} {\bf 591}, 255 (2004); 
\emph{Phys.~Lett.~B} {\bf 591}, 267 (2004).
\bibitem{bagan}
E.~Bagan, P.~Ball, and P.~Godzinsky, 
\emph{Phys.~Lett.~B} {\bf 301}, 249 (1993).
\bibitem{fulvia}
P.~Colangelo, F.~De Fazio, and N.~Paver, 
\emph{Phys.~Rev.~D} {\bf 58}, 116005 (1998).


\bibitem{Weinberg}
S.~Weinberg,
\emph{Phys.\ Rev.\ Lett.~}{\bf 110}, 261601 (2013). 

\bibitem{jaffe}
R.~L.~Jaffe, 
\emph{Nucl.~Phys.~A} {\bf 804}, 25 (2008).  

\bibitem{narison}
F.~Fanomezana, S.~Narison, and A.~Rabemananjara, 
arXiv:1409.8591. 

\bibitem{nielsen2010}
M.~Nielsen, F.~S.~Navarra, and S.-H.~Lee, 
\emph{Phys.~Rept.~} {\bf 497}, 41 (2010). 

\bibitem{nielsen2014}
F.~S.~Navarra and M.~Nielsen, 
\emph{Int.~J.~Mod.~Phys.~Lett.~A} {\bf 29}, 1430005 (2014).

\bibitem{wang}
Z.-G.~Wang and T.~Huang, 
\emph{Nucl.~Phys.~A} {\bf 930}, 63 (2014). 

\bibitem{erler}
J.~Erler, 
these proceedings, arXiv:1412.4435.

\bibitem{becirevic2}
D.~Becirevic, G.~Duplancic, B.~Klajn, B.~Melic, and F.~Sanfilippo, 
\emph{Nucl.~Phys.~B} {\bf 883}, 306 (2014). 

\bibitem{anikin}	
I.~V.~Anikin, V.~M.~Braun, and N.~Offen, 
\emph{Phys.~Rev.~D} {\bf 88}, 114021 (2013). 

\bibitem{aliev1}
T.~M.~Aliev, K.~Azizi, and M.~Savci, 
\emph{J.~Phys.~G} {\bf 41}, 065003 (2014). 
\bibitem{aliev2}
T.~M.~Aliev and M.~Savci, 
\emph{Phys.~Rev.~D} {\bf 89}, 053003 (2014). 

\bibitem{altug}	
A.~Kucukarslan, U.~Ozdem, A.~Ozpineci, 
\emph{Phys.~Rev.~D} {\bf 90}, 054002 (2014). 

\bibitem{bracco}
B.~Rodrigues, M.~E.~Bracco, M.~Chiapparini, and A.~Cerqueira, 
arXiv:1501.03088
\end{thebibliography}

\end{document}